\documentclass[%
 aps,prl,twocolumn,superscriptaddress,letterpaper,
 sd,%
 amsmath,amssymb,
reprint,%
]{revtex4-1}

\usepackage{soul, color}
\soulregister\ref{7}  
\soulregister\cite{7} 
\soulregister\onlinecite{7} 
\usepackage{graphicx}
\usepackage{dcolumn}
\usepackage{bm}

\begin{document}

\title[Broadband Angular Selectivity of Light at the Nanoscale: Progress, Applications and Outlook]{Broadband Angular Selectivity of Light at the Nanoscale: Progress, Applications and Outlook}

\author{Yichen Shen}
\email{ycshen@mit.edu}
\affiliation{Department of Physics, Massachusetts Institute of Technology, Cambridge, MA 02139, USA}
\affiliation{Research Laboratory for Electronics, Massachusetts Institute of Technology,Cambridge, MA 02139, USA}

\author{Chia Wei Hsu}%
\affiliation{Department of Physics, Massachusetts Institute of Technology, Cambridge, MA 02139, USA}
\affiliation{Department of Applied Physics, Yale University, New Haven, CT 06520, USA}

\author{Yi Xiang Yeng}
\affiliation{Research Laboratory for Electronics, Massachusetts Institute of Technology,Cambridge, MA 02139, USA}

\author{John D Joannopoulos}
\affiliation{Department of Physics, Massachusetts Institute of Technology, Cambridge, MA 02139, USA}
\affiliation{Research Laboratory for Electronics, Massachusetts Institute of Technology,Cambridge, MA 02139, USA}

\author{Marin Solja\v{c}i\'{c}}
\affiliation{Department of Physics, Massachusetts Institute of Technology, Cambridge, MA 02139, USA}
\affiliation{Research Laboratory for Electronics, Massachusetts Institute of Technology,Cambridge, MA 02139, USA}

\date{\today}

\begin{abstract}
Humankind has long endeavored to control the propagation direction of light. Since time immemorial, shades, lenses and mirrors have been used to control the flow of light. In modern society, with the rapid development of nanotechnology, the control of light is moving toward devices at micrometer and even nanometer scales. At such scales, traditional devices based on geometrical optics reach their fundamental diffraction limits and cease to work. Nano photonics, on the other hand, has attracted wide attention from researchers, especially in the last decade, due to its ability to manipulate light at the nanoscale. This review focuses on the nano photonics systems that aim to select light based on its propagation direction. In the first half of this review, we survey the literature and the current state of the art focused on enabling optical broadband angular selectivity. The mechanisms we review can be classified into three main categories: (i) microscale geometrical optics, (ii) multilayer birefringent materials and (iii) Brewster modes in plasmonic systems, photonic crystals and metamaterials. In the second half, we present two categories of potential applications for broadband angularly selective systems. The first category aims at enhancing the efficiency of solar energy harvesting, through photovoltaic process or solar thermal process. The second category aims at enhancing light extracting efficiency and detection sensitivity. Finally, we discuss the most prominent challenges in broadband angular selectivity and some prospects on how to solve these challenges.
\end{abstract}

\maketitle

\tableofcontents

\section{Introduction}
Following the invention of lasers in the 1960s, photonics has seen enormous advances in its impact on almost every discipline from everyday life to the most advanced science. The development of photonics technology has greatly enhanced our abilities in generation, emission, transmission and modulation of light. Traditional optical devices based on geometrical optics such as lenses, mirrors and fibers have done an excellent job of manipulating light in the past decades. However, just as electronics has seen a dramatic miniaturization over the last 40 years, which enabled data density to be doubled every 18 months (the so-called Moore's law), the feature sizes of photonic devices have rapidly decreased too. Unfortunately, such a trend has unavoidably slowed down because we are already at the boundary where geometrical optics breaks down (feature size $\approx$ $10\mu m$). In order to catch up with the density of electronic devices (feature size $\approx100nm$), the need to manipulate light at the micro-scale or even nano-scale is increasingly desirable for the optical community. 

\begin{figure}[htbp]
\begin{center}
\includegraphics[width=3.2in]{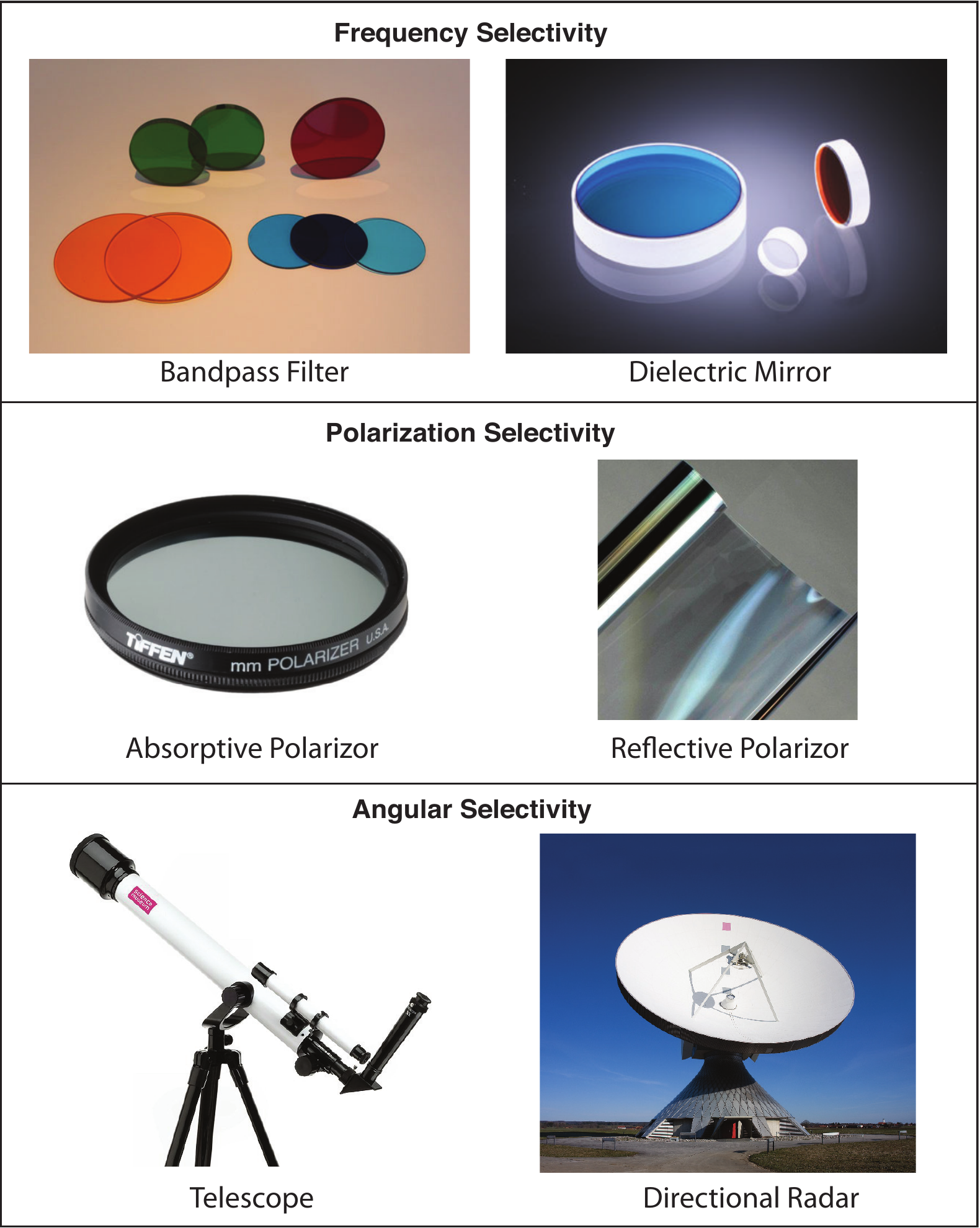}
\caption{\textbf{Conventional light selective devices.} Frequency selectivity can be achieved by bandpass filters \cite{bandpass_filter} or dielectric mirrors \cite{dielectric_mirror} (top panel); Polarization Selectivity can be achieved by wire grid polarizer, polaroid (polyvinyl alcohol polymer impregnated with iodine), or multilayer materials (3M Inc.) \cite{Weber31032000}; Traditional angularly selective devices are mainly based on geometric optics, such as telescopes \cite{telescope_figure} and radars \cite{Alemania_radar}.}
\label{fig:selectivity_compare}
\end{center}
\end{figure}

A monochromatic electromagnetic plane wave is characterized by three fundamental properties: frequency, polarization and propagation direction. The ability to select light according to each of these separate properties would be an essential step in achieving control over light. Over the past several decades, tremendous progress has been made towards achieving frequency selectivity and polarization selectivity on a scale far smaller than what geometrical optics permits (Fig.~\ref{fig:selectivity_compare}). For example, frequency selectivity can be obtained by using dyes, surface resonances \cite{xu2010plasmonic,chen2010high,si2013reflective,shen2015structural}, or taking advantage of photonic bandgaps in photonic crystals \cite{PhysRevLett.58.2059,Yablonovitch:93, joannopoulos2011photonic, SGJ_JDJ_3Dbandgap, fink1998dielectric}. The ability to filter light by its colors has resulted in modern flat panel display technology. On the other hand, polarization selectivity is accomplished by means of ``wire grid'' polarizers \cite{wiregrid:1985} or by exploiting birefringent materials \cite{Archard1948,hecht2008optics}. The ability to filter light by its polarization has to a large extent enabled the 3D movie industry. 

Light selection based on the direction of propagation, on the other hand, has seen relatively slower progress. Since ancient times, people have used a long tube in front of the eyes to effectively ``block out'' light from unwanted directions, or used louvers as a type of curtain to block out sunlight. Until now, most of the angularly selective systems have been based on geometrical optics approaches, such as the lens array systems in telescopes and the parabolic mirror disc in radar (Fig.~\ref{fig:selectivity_compare}). These systems are usually bulky and expensive.  

In recent years, the development of nanophotonics has led to many new devices at the wavelength or even sub-wavelength scale. Methods based on diffraction and plasmonic resonances have been explored for the purpose of angularly selection \cite{Upping2010102,Schwartz:03}; however, these effects are generally narrowband due to the inherent resonant properties of these mechanisms. Truly nanoscale \textit{broadband} angular selectivity has long been a scientific and engineering challenge \cite{Schwartz:03,de2007colloquium,le2012broadband,PhysRevLett.106.123902,Shen28032014,PhysRevB.90.125422,akozbek2012experimental,argyropoulos2013broadband,argyropoulos2012matching}. 

In this review, we mainly focus on the topic of nanoscale broadband angular selectivity. We start by presenting the key theoretical concepts, experiments and proposals in this field, namely using micro-scale geometrical optics, birefringent materials, metallic gratings, photonic crystals and metamaterials. We then extend our discussion to potential applications and impact that broadband angularly selective material could bring us. Finally, we discuss outlook for further theoretical and technological advances.

\section{Progress}

\subsection{Microscale Geometrical Optics}
As has been discussed in the introduction, traditional broadband angularly selective systems are mainly based on geometrical optics. Recently, in order to reduce the system size, efforts have been made to push to the smallest scale limit of geometrical optics. In these efforts, shades, lenses and mirrors are made at the micrometer scale. At such length scales they are too small to be seen by human eyes while still big enough to function in the regime where laws of geometrical optics hold. 

\begin{figure}[htbp]
\begin{center}
\includegraphics[width=3.2in]{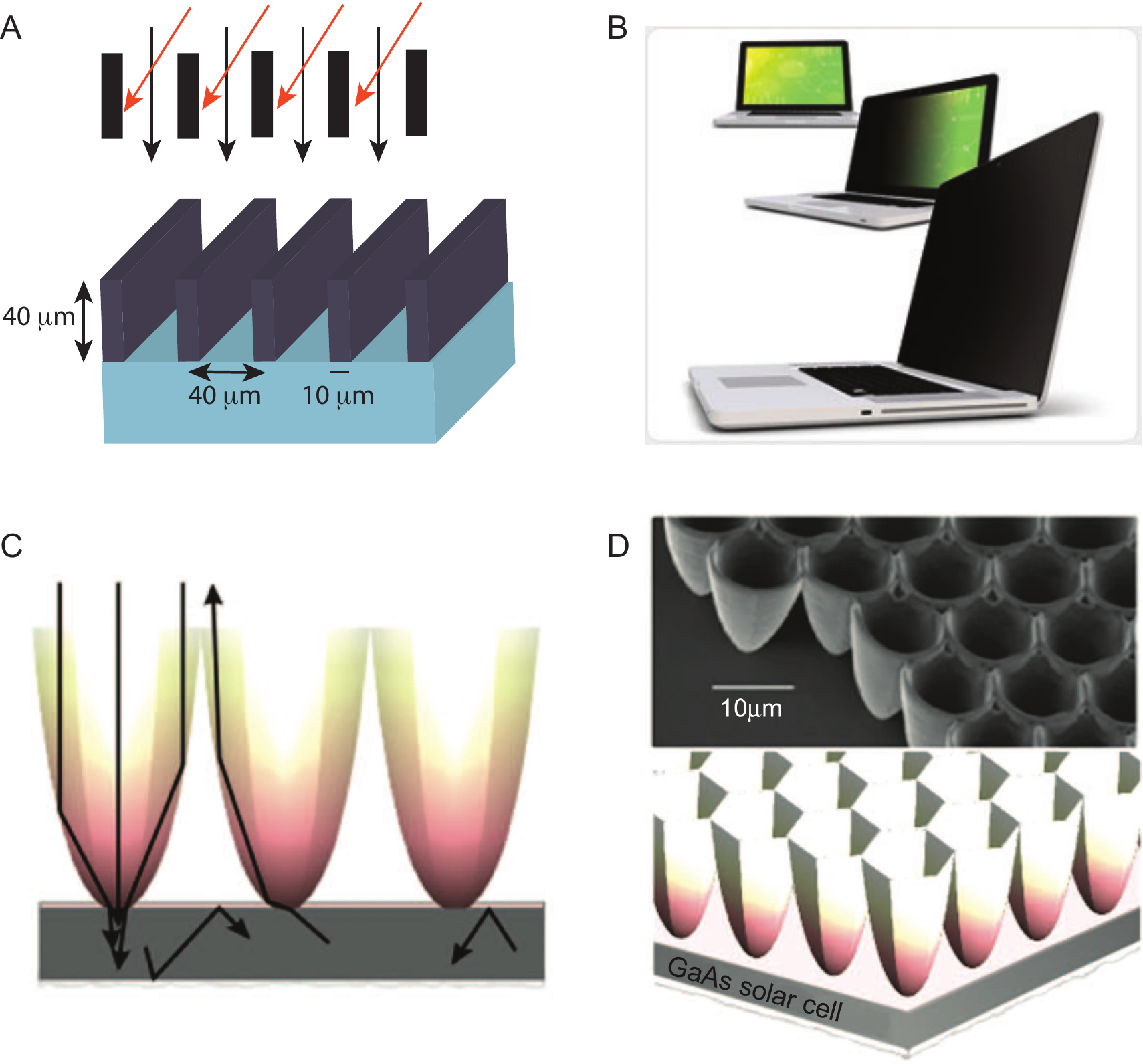}
\caption{\textbf{Angular Selectivity based on Microscale Geometrical Optics} (A) A schematic design of a micro-louvre angularly selective system \cite{privacyfilter:2005,lightcollimating:1988,lightcontrol:1983}. (B) The effect of the micro-louvre broadband angularly selective system as privacy protection. (C) Representative rays illustrate the function of the metallic coupler. (D) Schematic of metal array coupler on a solar cell with a randomizing back reflector and a scanning electron microscopy (SEM) of a structure fabricated in metal-coated resist via two-photon lithography.\cite{Atwater_angular2013,atwater2011microphotonic}}
\label{fig:microscale_optics}
\end{center}
\end{figure}

In mid 2000, motivated by the need to significantly narrow the usable viewing angle to provide viewing privacy, the broadband angularly selective effect was demonstrated using the micro-louvre approach by 3M Inc.\cite{privacyfilter:2005,lightcollimating:1988,lightcontrol:1983}. As shown in Fig.~\ref{fig:microscale_optics}A,B, their angularly selective system comprises of (i) a transparent polymeric material as the base sheet; and (ii) light directing elements comprising absorptive material, wherein each element served as light shade with width $w$, height $h$, and distance $a$ away from each other. In this micro-louvre design, $w$, $h$ and $a$ are all in the range from $5\mu m$ to $50\mu m$, so human eyes cannot see the micro-louvre components, but they are still much bigger than the wavelength of the visible light, so laws of geometrical optics still hold.

Recently, a reflective broadband angularly selective filter based on microscale geometrical optics was proposed and fabricated by Atwater and co-workers \cite{Atwater_angular2013, atwater2011microphotonic} (Fig.~\ref{fig:microscale_optics}C,D). Their simulation showed that the parabolic directors they designed with $22\mu m$ height and $10\mu m$ diameter exhibit a strong angular selectivity. Light with incident angle less than $5.6^{\circ}$ is able to funnel through the system, while light incident at larger angles is strongly reflected. In their work, the parabolic light director is fabricated using two-photon lithography, thin-film processing, and aperture formation by focused ion beam lithography \cite{atwater2011microphotonic}. The proposed system was later shown by the same group to have a potential (in theory) to increase the efficiency of solar cells \cite{Atwater_angular2013}. However, a successful experimental demonstration of this proposed angularly selective absorber has yet to be reported due to the difficulties in high-resolution large scale fabrication.

In general, broadband angularly selective systems based on micro-geometrical optics designs have enabled new opportunities in managing light's direction on small scales. However, the performance of systems based on such a mechanism will ultimately be limited by diffraction and scattering effects when the scales of interest approach the geometrical optics limit.

\subsection{Birefringent-Polarizer Systems}
Another method that has been proposed to achieve broadband angular selectivity is to use a combination of polarizers and birefringent films \cite{privacyfilter:2001,privacyfilter:2003,Mathew_angular2002}. This new approach is not based on geometrical optics. However, in reality, the birefringent material still needs to be many wavelengths thick in order to have enough phase retardation.

In a uniaxially birefringent film, if the extraordinary index (or the optical axis) lies in the plane of the film, it is said to be a half-wave plate if its thickness $d_1$ and birefringence $n_e-n_o$, are chosen such that:
\[(n_e-n_o)d=\lambda/2\]
where $\lambda$ is the wavelength of incident light. Half-waveplates have the property that when plane polarized light is incident in the direction such that the polarization vector makes an angle $\theta$ with the extraordinary axis of the film, then the plane of polarization will be rotated by $2\theta$ as the light passes through the plate. The half wave plate can be made broadband by stacking several different birefringent layers together to compensate for the frequency dispersion.

\begin{figure}[htbp]
\begin{center}
\includegraphics[width=3.2in]{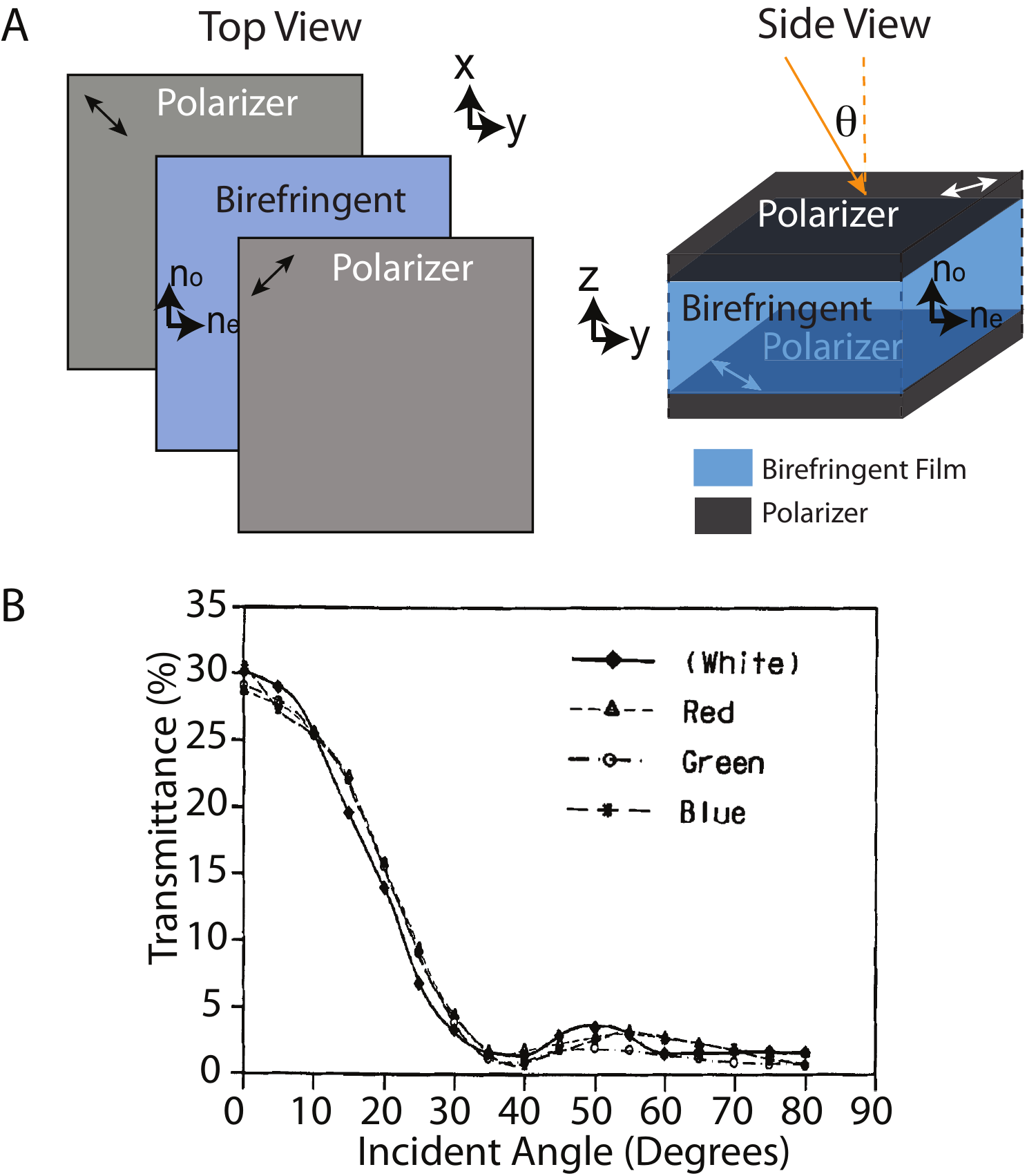}
\caption{\textbf{Birefringent Broadband Angularly Selective Film} (A) Schematic layout of a birefringent angularly selective filter. (B) Experimental measurement result of the transmission spectrum (random polarization) of the sample at different incident angles \cite{privacyfilter:2003}.}
\label{fig:birefringent}
\end{center}
\end{figure}

Fig.~\ref{fig:birefringent} illustrates the simplest design that operates with this principle. The design comprises a half wave plate sandwiched between two polarizers. The polarizing axes of the two polarizers are crossed at 90$^{\circ}$. Upon passing the first polarizer and the birefringent film, the polarization axis of the light is rotated appreciably. The rotation of the polarization axis is proportional to the distance that the light travels when passing through the birefringent film. For orthogonal light, the distance traversed in passing through the birefringent film is minimal and equal to the thickness $d$ of the film. For oblique light, the distance is greater than $d$ and depends upon the incident angle of the light. The degree of birefringence and the thickness of the birefringent film are chosen appropriately in relation to the optical and polarizing axes of the angularly selective film such that all orthogonal light has its polarizing axis rotated by $90^{\circ}$ after passing through the birefringent film and is thus transmitted through the second polarizer, while non-orthogonal light has its polarizing axis rotated by a different angle, and is hence substantially blocked by the second polarizer.

Fig.~\ref{fig:birefringent}B presents the experimental result of one such design \cite{privacyfilter:2003}; It is composed of two linear polarizing laminated films (LP). The birefringent material was 27 mil Cellulose Diacetate film (CDA). The CDA and LP were separated by a $1/16''$spacer with an open window aperture of $4.5''\times3''$. The optical axis of the CDA was oriented horizontally (along the x axis with z being orthogonal to the $xy$ plane of the film and x being the horizontal direction) with respect to the film axis and the LP was aligned with the polarizing axis oriented $\pm45^{\circ}$ to the optical axis of the CDA. As one can see, the combined material system has relatively high transparency for light at normal incidence, with an angularly selective window of around $30^{\circ}$.

Because the birefringence-polarizer angularly selective system is simple to fabricate and can be easily scaled up to a large area, it has found its most common application in privacy protection films. However, as this mechanism relies solely on the rotation of the light polarization direction through the birefringent film, the angular window of transparency is typically quite broad ($>20^{\circ}$). Furthermore, the incident light has to pass through two polarizers, and as a result, the peak transmissivity is heavily attenuated.

\subsection{Plasmonic Brewster Angle in Metallic Gratings}
Extraordinary optical transmission (EOT) \cite{ebbesen1998extraordinary,garcia2010light,de2007colloquium} is the phenomenon of greatly enhanced transmission of light incident at certain directions through a subwavelength aperture in an otherwise opaque metallic film which has been patterned with a regularly repeating periodic structure. Therefore, it is a natural candidate for providing angular selectivity. However, the traditional EOT is based on plasmonic resonances, so it inherently has a limited bandwidth of operation.

\begin{figure}[htbp]
\begin{center}
\includegraphics[width=3.2in]{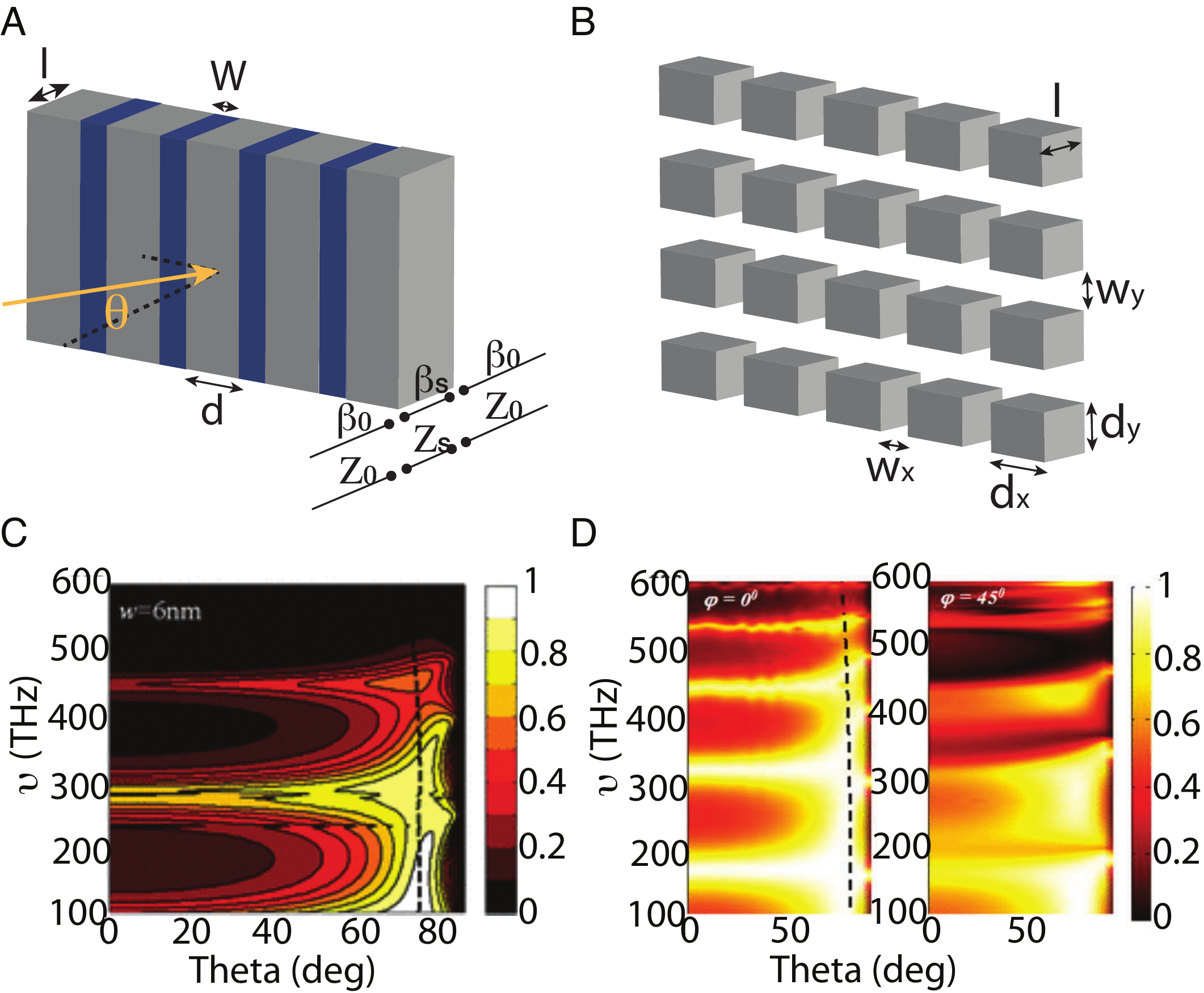}
\caption{{\bf Angular filter based on plasmonic Brewster angle in metallic gratings.}
(A) The geometry of the 1-dimensional grating (B) The geometry of the 2-dimensional grating. (C) Angular transmission spectra for a 1-dimensional grating with thickness $l=200$nm, period $d=96$nm, and slit width $w=6$nm. The dashed line indicates the plasmonic Brewster angle condition~\cite{PhysRevLett.106.123902}. (D) Angular transmission spectra for a 2-dimensional grating thickness $l=400$nm, period $d_x=d_y=100$nm, and slit width $w_x=6$nm and $w_y=12$nm. The dashed line indicates the plasmonic Brewster angle condition~\cite{le2012broadband}.}
\label{fig:fig2}
\end{center}
\end{figure}

Instead of relying on plasmonic resonances inside the metallic film, Al$\grave{u}$ \textit{et al} \cite{PhysRevLett.106.123902,akozbek2012experimental,argyropoulos2013broadband,argyropoulos2012matching} used a nonresonant Brewster-like effect based on impedance matching to achieve enhanced transmission through metallic gratings (Fig.~\ref{fig:fig2}). This mechanism provides enhanced transmission at a particular angle for transverse-magnetic (TM) polarization and is weakly dependent on the frequency.

To understand this plasmonic Brewster angle mechanism, consider the geometry of Fig.~\ref{fig:fig2}A: A metallic screen of thickness $l$, corrugated by slits of width $W$ and period $d$, is illuminated by a TM wave. The scattering from such a periodic structure can be modeled by using a transmission-line (TL) approach \cite{medina2008extraordinary}. The circuit analog is depicted at the bottom of Fig.~\ref{fig:fig2}A: the free space region is modeled as a semi-infinite transmission line; for a given angle of incidence $\theta$ with respect to the $z$ axis, the effective vacuum wavenumber is $\beta_0=k_0\cos\theta$, the characteristic impedance per unit length of vacuum can be calculated by
\begin{equation}
Z_0=\frac{V_0}{I_0}=\frac{\int_0^{d}E_xdx}{H_y}=\frac{|E|d\cos\theta}{|E|/\eta_0}=\eta_0d\cos\theta.
\label{eqn:z0}
\end{equation}
Where $\eta_0=\sqrt{\mu_0/\epsilon_0}$ is the vacuum impedance. Inside each slit, modal propagation does not depend on the incidence angle. When the slit is significantly narrower than the periodicity and the wavelength, that is, only the dominant TM mode propagates inside the slit and only the fundamental diffraction order radiates in free space, i.e., $W\ll d<\lambda_0=2\pi/k_0$, and the characteristic impedance per unit length $Z_s$ satisfies the equation\cite{alu2006optical}
\begin{equation}
Z_s=W\beta_s/(\omega\epsilon_0),
\label{eqn:zs}
\end{equation}
where the wavenumber $\beta_s$ satisfies the equation
\[\tanh\left[\sqrt{\beta_s^2-k_0^2}W/2\right]\sqrt{\beta_s^2-k_0^2}=-\sqrt{\beta_s^2-k_0^2\epsilon_m}/\epsilon_m.\]
From Eqn~\ref{eqn:z0} and Eqn.~\ref{eqn:zs}, we observe that the characteristic impedances of air and slit are not equal in general, but as we change the incident angle $\theta$, there can exist an angle $\theta_B$, or the so-called plasmonic Brewster angle, such that the two impedances are equal $Z_0(\theta_B)=Z_s$. The condition for such impedance matching is:
\begin{equation}
\cos\theta_B=(\beta_sW)/(k_0d).
\label{eqn:impedance_match}
\end{equation}
This impedance matching is independent of the grating thickness $l$ and only depends weakly on the frequency $\omega$, as shown in Fig.~\ref{fig:fig2}C. In fact, the angularly selective working frequency range extends from zero frequency (DC) to the breakdown of this model $d\approx\lambda_0$. 

The mechanism described above using 1D metallic gratings only works for light incident in the plane perpendicular to the slit direction. In a later paper \cite{le2012broadband}, the same group led by Al$\grave{u}$ demonstrated that such a phenomenon, representing the equivalent of Brewster transmission for plasmonic screens, can also occur in 2D metallic gratings of various structural forms and shapes, and that it may be made less sensitive to the azimuthal angle $\phi$ (Fig.~\ref{fig:fig2}B,D). 

Although the experimental realization has been demonstrated in the microwave regime, it is challenging to realize such plasmonic Brewster mode in the visible wavelength regime, mainly due to the metallic loss in the visible regime, and the challenging fabrication requirement (needed resolution $<5$nm). However, the Brewster mode idea proposed here has inspired subsequent research that brings broadband angular selectivity to new levels of performance.

\subsection{Optical Brewster Modes in One-Dimensional Photonic Crystals}
Photonic crystals (PhCs) are periodic structures \cite{joannopoulos2011photonic} in which the index of refraction varies between high-index and low-index, with a periodicity comparable to the wavelength. Such an environment presents to photons what the periodic atomic potential of a semiconductor presents to electrons. In particular, just as a periodic atomic potential will open up a band gap inside the crystal where no electron is allowed to propagate, under the right conditions, a PhC can exhibit a complete photonic band gap: a range of frequencies in in which light is forbidden to propagate inside the material. 

\begin{figure*}[htbp]
\begin{center}
\includegraphics[width=6in]{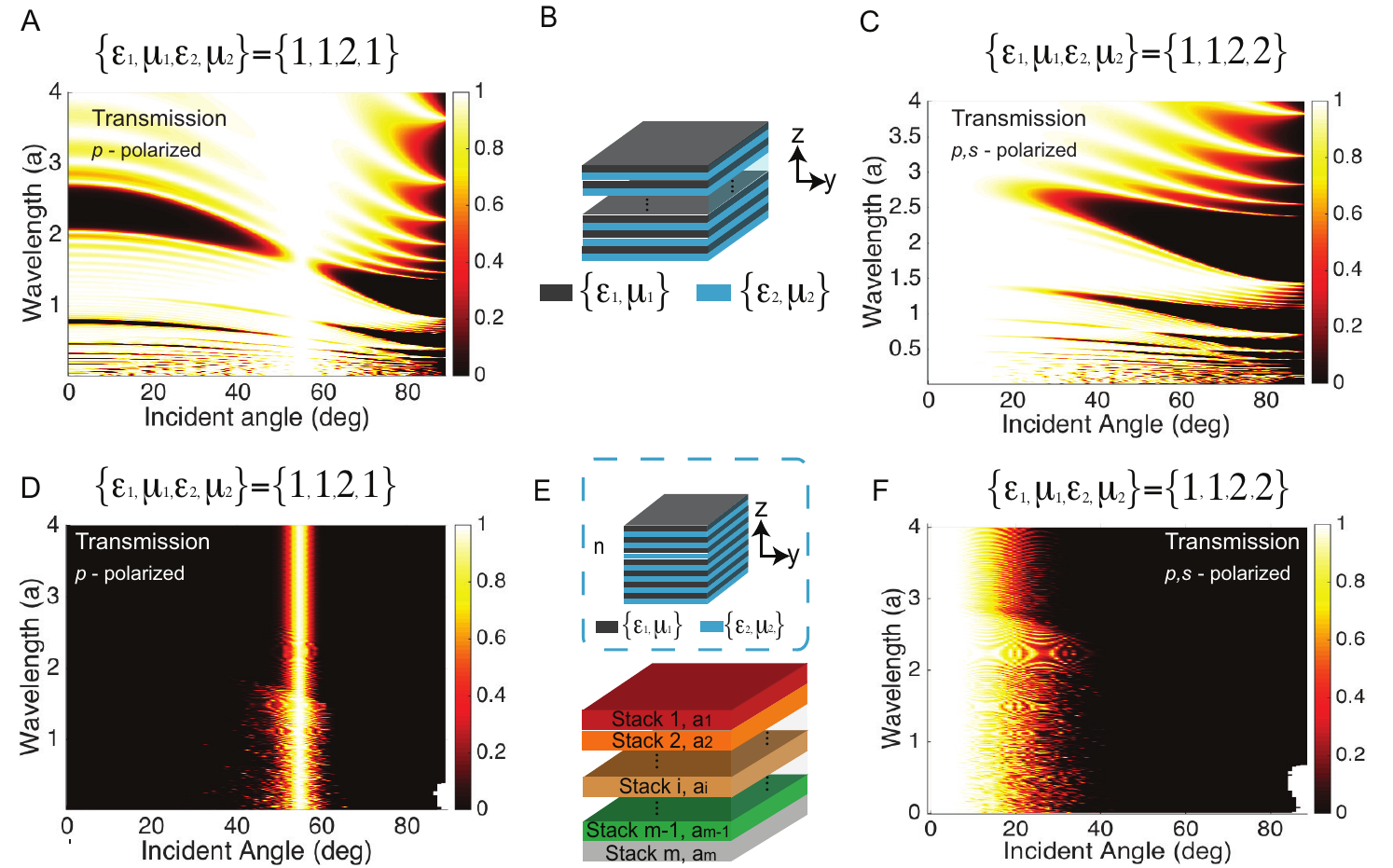}
\caption{{\bf Design of angular filter based on Brewster modes in 1D photonic crystals.}
(A) $p$-polarized transmission spectrum of a quarter-wave stack with 20 alternative layers of two materials having $\epsilon_1=1$ and $\epsilon_2=2$ respectively. (B) Schematic layout of a simple quarter-wave stack. (C) The same plot as in part (A), but with $\epsilon_1=\mu_1=1$, $\epsilon_2=\mu_2=2$, and for both polarizations. (D) $p$-polarized transmission spectrum of 50 stacks of quarter-wave stacks at various periodicities. Each quarter-wave stack consists of 10 bi-layers of $\{\epsilon_1=1$,$\epsilon_2=2\}$ materials. The periodicities of these quarter-wave stacks form a geometric series $a_i=a_0r^{i-1}$ with $r=1.0212$, where $a_i$ is the periodicity of $i^{th}$ stack. (E) Schematic layout of the heterostructure stacking mechanism (F) $p,s$-polarized transmission spectrum for a structure that has the same number of stacks and layers per stack as in part (D), but with different material properties: $\epsilon_1=\mu_1=1$, $\epsilon_2=\mu_2=2$. 
}
\label{fig:phc_angle}
\end{center}
\end{figure*}

\begin{figure}[htbp]
\begin{center}
\includegraphics[width=3.2in]{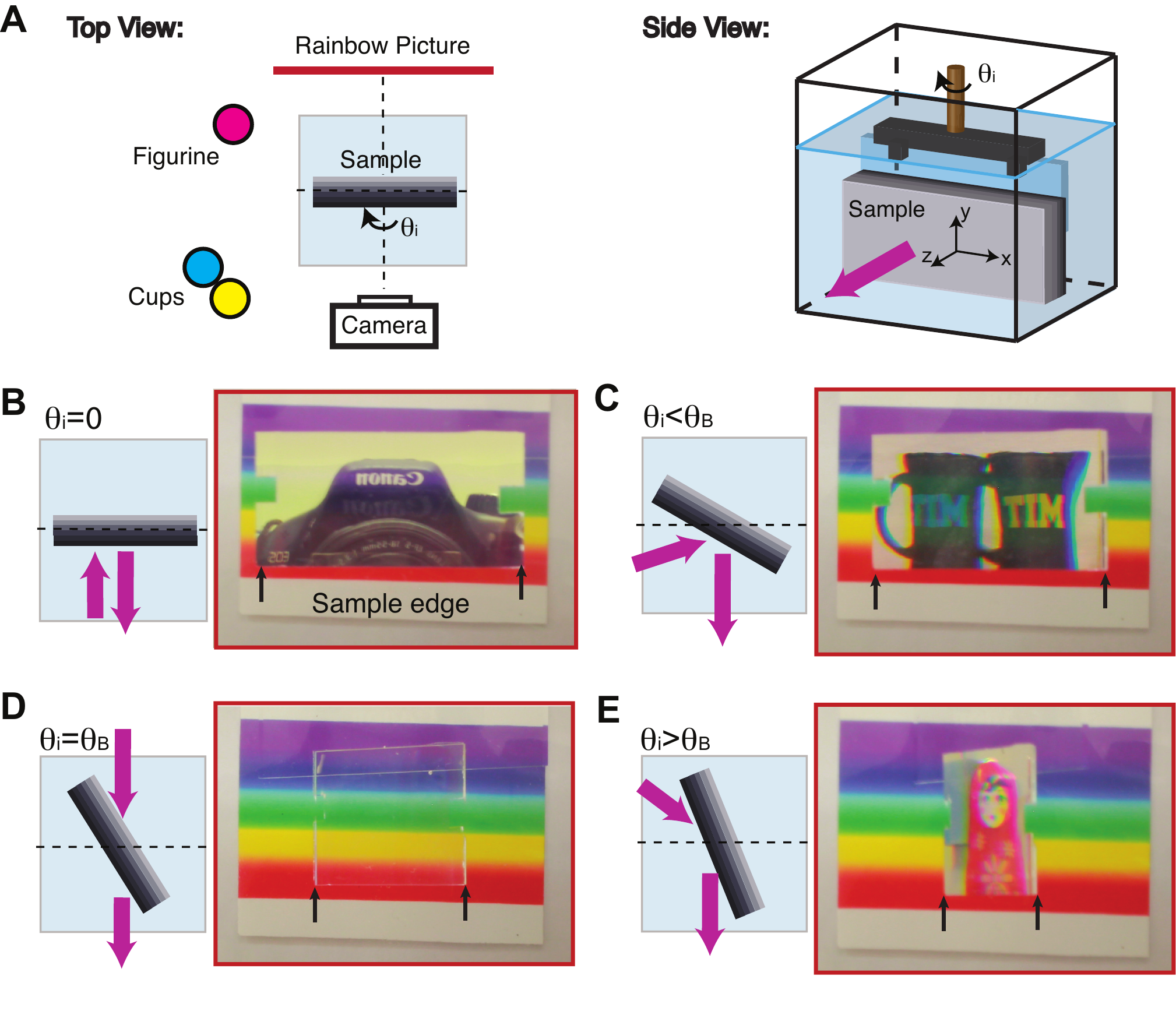}
\caption{\textbf{Experimental realization of the 1D-PhC angular filter.} (A) Schematic layout of the experimental setup. The system is immersed in a liquid that is index matched to $\epsilon_1=\epsilon_{SiO_2}=2.18$. (B)Normal incident angle setup. The sample behaves like a mirror and reflects the image of the camera. (C) $\theta_i=30^{\circ}$ setup. The sample behaves like a mirror and reflects the image of MIT cups in the lab. (D) $\theta_i=\theta_B=55^{\circ}$ setup. The sample becomes transparent for the entire visible regime for $p-$polarized light. (E) $\theta_i=70^{\circ}$ setup. The sample behaves as a mirror and reflects the figurine placed at the corner of the table. In B-E a polarizer is installed on the camera so it detects only $p$-polarized light~\cite{Shen28032014}}
\label{fig:brewster_angular}
\end{center}
\end{figure}

This unique property of blocking light using photonic band gaps leads to the following intriguing possibility: imagine if one could secure a direction in a one dimensional (1-D) PhC such that incident light always passes through (regardless of frequencies and the thickness of each layer), and could design the thickness of each layer of the 1-D PhC such that light propagating in other directions is all blocked by the photonic band gaps; then the resulting material system would potentially have broadband angularly selective behavior.

Such an idea was proposed and realized in a recent work by Shen \textit{et al.} \cite{Shen28032014}. The design principle rests on three different features of a 1-D PhC: (i) $p$-polarized light transmits without any reflection at the Brewster angle of each interface, (ii) the existence of band gaps that prevent light propagation for given frequency ranges, and (iii) the band-gap-broadening effect of heterostructures (Fig.~\ref{fig:phc_angle}D,F).

Consider the light reflection at the interface between two dielectric materials with permittivities $\epsilon_1$, $\epsilon_2$ and permeabilities $\mu_1$, $\mu_2$ respectively. The off-axis reflectivity can be calculated directly from the generalized Fresnel equations \textit{\cite{hecht2008optics}}:
\begin{equation}
\left(\frac{E_{r}}{E_i} \right )_\perp=\frac{\frac{1}{Z_i}\cos\theta_i-\frac{1}{Z_t}\cos\theta_t}{\frac{1}{Z_i}\cos\theta_i+\frac{1}{Z_t}\cos\theta_t}
\label{eqn:Fresnel1}
\end{equation}
and
\begin{equation}
\left(\frac{E_{r}}{E_i} \right )_\parallel =\frac{\frac{1}{Z_t}\cos\theta_i-\frac{1}{Z_i}\cos\theta_t}{\frac{1}{Z_i}\cos\theta_i+\frac{1}{Z_t}\cos\theta_t},
\label{eqn:Fresnel2}
\end{equation}
where the impedance $Z$ is defined as $Z=\sqrt{\frac{\mu}{\epsilon}}$, the subscripts $i$ and $r$ denote incident light and reflected light, respectively, and the subscripts $\perp$ and $\parallel$ indicate the direction of the electric field $\overrightarrow{E}$ with respect to the plane of incidence. At certain incident angle $\theta_B$ (so-called Brewster angle) when the numerators in Eqn.~\ref{eqn:Fresnel1} and Eqn.~\ref{eqn:Fresnel2} become zero, there is no reflection at the interface.
Notice for the 1D PhC, the light propagating at the Brewster angle from $\epsilon_1$ material to $\epsilon_2$ material automatically satisfies the Brewster condition from $\epsilon_2$ material to $\epsilon_1$.

Next, consider a simple quarter-wave stack consisting of these two materials (Fig.~\ref{fig:phc_angle}B). In such a system, monochromatic plane waves with frequency $\omega$ cannot propagate in certain directions due to destructive interference, or so-called photonic band-gaps (dark region in Fig.~\ref{fig:phc_angle}A,C). However, when the light is propagating at the Brewster angle (55$^{\circ}$ in Fig.~\ref{fig:phc_angle}A and 0$^{\circ}$ in Fig.~\ref{fig:phc_angle}C), since it has no reflections at any interface, destructive interference does not happen, and the light always goes through.The transmission spectrum of such a quarter-wave stack is shown in Figs.~\ref{fig:phc_angle}A,C. Furthermore, from Figs.~\ref{fig:phc_angle}A and C, we can see that a single quarter-wave stack is not sufficient to provide broadband angular selectivity. What one needs is a bigger photonic band gap to cover the entire spectrum at off-Brewster angles. The enlargement of photonic band gap  is achieved by stacking togther quarter-wave stacks with various periodicities \cite{perilloux2002thin,Zhang2000,Xin2002heterostructure} (Figs.~\ref{fig:phc_angle}D,F). 

The experimental demonstration of this concept was shown by Shen \textit{et al} \cite{Shen28032014} (Fig.~\ref{fig:brewster_angular}). The fabricated sample consists of SiO$_{2}$ and Ta$_2$O$_5$. The sample is transparent (up to 98\%) to $p$-polarized incident light at $\theta_B=55^{\circ}$ (Fig.~\ref{fig:brewster_angular}D); the angular window of transparency is about 8$^{\circ}$. The sample behaves like a mirror at all other incident angles over the entire visible spectrum (Fig.~\ref{fig:brewster_angular}B,C,E). 

\begin{figure}[htbp]
\begin{center}
\includegraphics[width=3.5in]{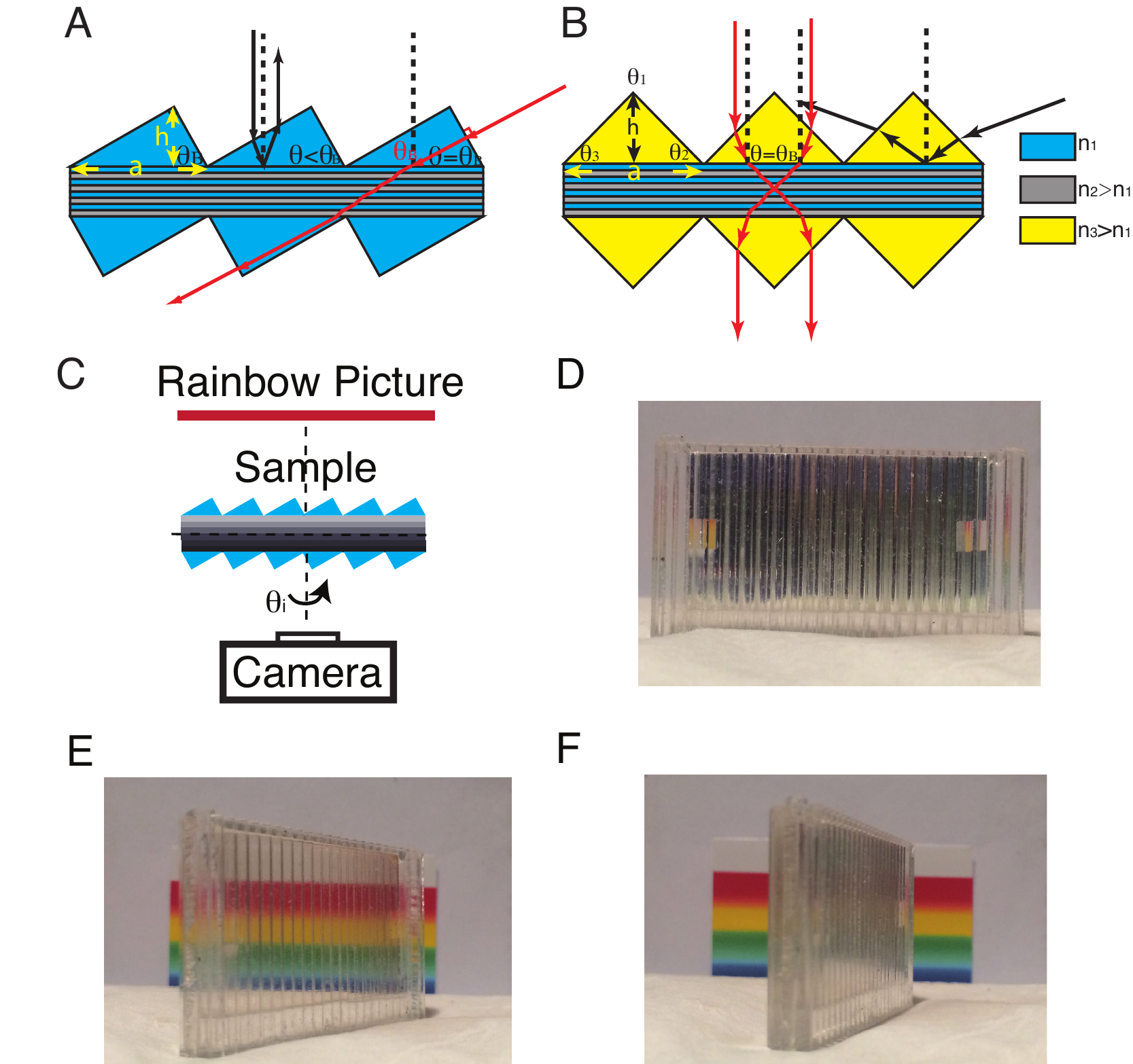}
\caption{\textbf{Illustration of optical angularly selective filter in air using prisms} (A) Oblique incidence prism design (B) Normal incidence prism design. The red beams represent the light rays coming in at the right direction so the prisms can couple the light into the Brewster mode inside the PhC slab. The black beams represent the light coming in from other directions so they cannot be coupled into the Brewster mode inside the PhC slab; hence, they are reflected back into the prisms. (C) Schematic view of the system setup. (D) Normal incidence angle setup: the sample is reflective. (E) $\theta=55^{\circ}$ setup: the sample becomes transparent. (F) $\theta=75^{\circ}$ setup: the sample become reflective again.}
\label{fig:brewster_prism}
\end{center}
\end{figure}

In the original experiment, the sample had to be immersed in an index-matched liquid. In-air operation requires ultra-low index materials (such as aerogel \cite{aegerter2011aerogels}) as one of the composition layers, which makes it hard to fabricate. Furthermore, the Brewster angle at the interface of two dielectric media (in the lower index isotropic material) is always larger than 45$^{\circ}$. An implementation introduced later by the same group used macroscopic prisms to improve the previous design, such that the device can be directly used in air without using ultra-low index materials. In addition, the angle of high transparency can be adjusted easily over a broad range, including normal incidence \cite{air_compatible} (Fig.~\ref{fig:brewster_prism}). Notice for normal angular-window design shown in Fig.~\ref{fig:brewster_prism}B, the image is not preserved. However, the image can be preserved if one adds another layer of angular filter, as described in Ref.~\cite{air_compatible}.

\subsection{Brewster Modes in Metamaterials}
Metamaterials are artificially created materials engineered to have properties that may not be found in nature. They are assemblies of building blocks, each consisting of several different conventional material types such as metals or plastics. The building blocks are usually arranged in periodic patterns with a deep-subwavelength periodicity. Metamaterials gain their properties not from their fundamental composition (crystal structure, electron distribution, etc.), but from their artificially-designed microscopic building blocks. Their precise shape, geometry, size, orientation and arrangement can affect the waves of light or sound in an unconventional manner, creating material properties that cannot be achieved with conventional materials. 

\begin{figure}[htbp]
\begin{center}
\includegraphics[width=3.5in]{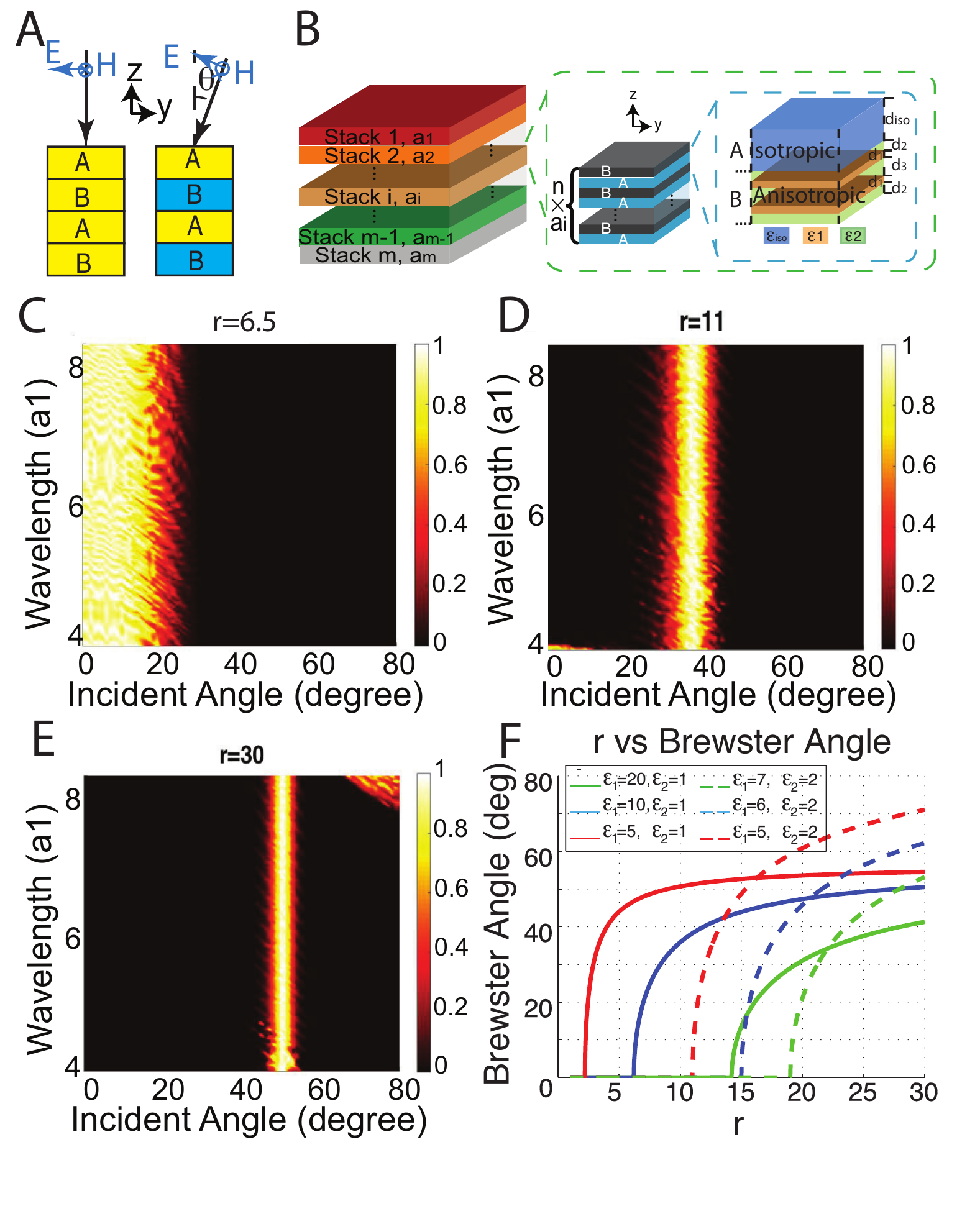}
\caption{{\bf Angular filter based on Brewster angle in metamaterials.}
(A) Schematic illustration of a stack of isotropic-anisotropic photonic crystals. Layer A is an isotropic medium; layer B is an effective anisotropic medium consisting of two different isotropic media with dielectric constants $\epsilon_1$ and $\epsilon_2$. The thickness ratio between these two materials is set to be $r$. (B) Effective index for $p$-polarized light in an isotropic A - anisotropic B multilayer system. Left panel: all the layers have the same index. Right panel: the change in incident angle leads to a change in the observed index of the anisotropic material layers B. (C-E) Transmission spectrum for the metamaterial photonic crystal design. The materials and structures are illustrated in subfigure A, with $n = m = 30$, $\epsilon_1=10$, $\epsilon_2=1$, and $\epsilon_{iso}=2.25$, but different $r$. (C) $r=6.5$ and $\theta_B=0^{\circ}$ (D) $r = 9$ and $\theta_B=24^{\circ}$. (E) $r=11$ and $\theta_B=38^{\circ}$. (E) $r = 30$ and $\theta_B = 50^{\circ}$. (F) Dependence of the Brewster angle (coupled in from air) on $r$ for various values of $\epsilon_1$ and $\epsilon_2$. Solid and dashed lines, correspond, respectively, to $\epsilon_1$ for air and $\epsilon_2$ for PDMS \cite{PhysRevB.90.125422}.}
\label{fig:metamaterial}
\end{center}
\end{figure}

In 2011, Hamam \textit{et al} \cite{PhysRevA.83.035806} pointed out that an angular photonic band gap can exist within anisotropic material systems (Fig.~\ref{fig:metamaterial}A). Inspired by this and building upon their earlier work of using Optical Brewster Modes in 1-D PhC \cite{Shen28032014}, Shen \textit{et al} proposed a new design that replaces one of the isotropic materials in conventional 1-D PhC by effective-anisotropic-metamaterials. In this way one can in principle achieve a broadband angularly selective behavior at \textit{arbitrary} incident angles \cite{PhysRevB.90.125422}.

To understand this efect, one can follow the same line of argument as in the previous section, but this time consider the light reflection at the interface between an isotropic medium with dielectric constant $\epsilon_{iso}$ and an anisotropic medium with dielectric constant $\{\epsilon_x,\epsilon_y,\epsilon_z\}$. The reflectivity of $p$-polarized light with a propagating angle $\theta_{i}$ (defined in the isotropic material) at an isotropic-anisotropic interface is \cite{azzam1977}
\begin{equation}
R_p=\left|\frac{n_xn_z\cos{\theta_{i}}-n_{iso}(n_z^2-n_{\rm iso}^2\sin{\theta_{i}}^2)^{\frac{1}{2}}}{n_xn_z\cos{\theta_{i}}+n_{\rm iso}(n_z^2-n_{\rm iso}^2\sin{\theta_{i}}^2)^{\frac{1}{2}}}\right|^2
\label{eqn:iso-ani}
\end{equation}
where $n_x=n_y$ and $n_z$ are the refractive indices of the anisotropic material at the ordinary and extraordinary axes, respectively, and $n_{\rm iso}$ is the refractive index of the isotropic material.

Therefore, the Brewster angle, $\theta_i=\theta_B$, can be calculated by setting $R_p=0$, giving:
\begin{equation}
\tan{\theta_B}=\sqrt{(\frac{\epsilon_z}{\epsilon_{\rm iso}})\left[\frac{\frac{\epsilon_x}{\epsilon_{\rm iso}}-1}{\frac{\epsilon_z}{\epsilon_{\rm iso}}-1}\right]}
\label{eqn:brewster_meta}
\end{equation}

Recent progress by 3M Inc. has shown that fabricating such isotropic-anisotropic multilayer PhC might be possible \cite{Weber31032000}, even for visible light. Furthermore, following this idea, Shen \textit{et al} \cite{PhysRevB.90.125422} introduced tunability of the dielectric constant of the anisotropic layer by using metamaterial to replace the anisotropic layers in Fig.~\ref{fig:metamaterial}A. As shown in Fig.~\ref{fig:metamaterial}B, each metamaterial layer consists of several high-index ($\epsilon_1=10$) and low-index ($\epsilon_2=\epsilon_{air}=1$) isotropic material layers. When the combined high-index-low-index two-layer unit is sufficiently thin compared to the wavelength (optically thin), the whole system can be treated as a single anisotropic medium with the \textit{effective} dielectric permittivity tensor $\{\epsilon_x,\epsilon_y,\epsilon_z\}$ \cite{Bergman1978377}:
\begin{align}
\label{eqn:1}
\epsilon_x&=\epsilon_y=\frac{\epsilon_1+r\epsilon_2}{1+r} \\
\label{eqn:2}
\frac{1}{\epsilon_z}&=\frac{1}{1+r}\left(\frac{1}{\epsilon_1}+\frac{r}{\epsilon_2}\right)
\end{align}
where $r$ is the ratio of the thickness of the two materials $\epsilon_1$ and $\epsilon_2$: $r=d_2/d_1$. Substituting Eqn.~\ref{eqn:1} and Eqn.~\ref{eqn:2} into Eqn.~\ref{eqn:brewster_meta} gives the dependence of Brewster angle $\theta_B$ on $r$:
\begin{equation}
\theta_B(r)=\arctan\left[\sqrt{\frac{\epsilon_1'\epsilon_2'(\epsilon_1'+r\epsilon_2'-1-r)}{(1+r)\epsilon_1'\epsilon_2'-\epsilon_2'-\epsilon_1'r}}\right ]
\label{eqn:brewster_final}
\end{equation}
where $\epsilon_1'=\epsilon_1/\epsilon_{\rm iso}$, and $\epsilon_2'=\epsilon_2/\epsilon_{\rm iso}$.

Eqn.~\ref{eqn:brewster_final} shows that it is possible to adjust the Brewster angle by changing the ratio $r=d_1/d_2$, which is confirmed by transfer matrix calculation of the transmission spectrum of an isotropic-anisotropic PhC with $n=m=30$ but different $r$ (Fig.~\ref{fig:metamaterial}C-E). The dependence of the Brewster angle on $r$ is presented in Fig.~\ref{fig:metamaterial}F. At small $r$, there is either no Brewster angle or the light cannot be coupled into the Brewster angle from air. As $r$ gets larger, a rapid increase in the Brewster angle is observed, which eventually plateaus, approaching the isotropic-isotropic limit, $\theta_B=\arctan\sqrt{\epsilon_2/\epsilon_{\rm iso}}$ \cite{hecht2008optics}. Note that if $\epsilon_2$ is some soft elastic material (such as PDMS or air), one can simply vary $r$ by changing the distance $d_2$ in real time, and hence varying the Brewster angle accordingly. Such \textit{tunability} of the Brewster angle does not exist in conventional (non-metamaterial) isotropic-isotropic or isotropic-anisotropic photonic crystals, where the Brewster angle depends solely on the materials' dielectric properties.

\section{Applications}
\subsection{Solar Energy Harvesting}

One major application for the reflective angularly selective material system is in solar energy harvesting, such as in solar cells and solar thermal systems.

Current solar energy harvesting technologies for electricity production can be categorized into two main conversion processes: direct conversion and indirect conversion. Direct conversion is also known as photovoltaic conversion (solar cells). In an ideal photovoltaic conversion, each photon absorbed by a solar-active material (typically silicon) causes an electron to jump from the valence band to the conduction band and create a hole in the valence band, leading to an electric current pulse \cite{einstein1965}. On the other hand, indirect conversion, or solar thermal conversion, corresponds to conversion of solar energy into heat by collection in a solar absorber and subsequent conversion of this heat into work by means of a thermal engine \cite{solarthermal2013}, a thermal photovoltaic device\cite{Yeng23012012,lenert2014,Veronika_superlattice_2014}, or a thermal electrical device\cite{disalvo1999thermoelectric,venkatasubramanian2001thin}.

The total efficiency of a solar cell (SC) or solar thermal (ST) system can be calculated by:
\begin{equation}
\mu_{total}=\mu_{I}\cdot\mu_{SC/ST}
\end{equation}
where $\mu_{I}$ represents the efficiency of light trapping, which is the percentage of solar energy that is trapped in the semiconductor layer (for solar cells) or absorber (for solar thermal systems), while $\mu_{SC/ST}$ represents the efficiency of the actual photo-electric conversion in solar cells, or thermal-electric conversion in solar thermal systems. Since $\mu_{SC/ST}$ is mainly limited by the inherent material properties such as the Shockley-Queisser limit \cite{shockley1961} (for solar cells) and Carnot efficiency (for solar thermal systems), here we focus on applying angularly selective surface to trap light and optimize $\mu_I$.

The efficiency of light trapping $\mu_I$ is highly sensitive to the absorber's emittance and can be calculated by:
\begin{equation}
\mu_{I}=\frac{P_{\rm absorbed}}{P_{\rm inc}}=\frac{J_{s}-J_{e}-J_{r}-J_{o}}{J_{s}}=1-\frac{J_{e}}{J_{s}}-\frac{J_r}{J_s}-\frac{J_o}{J_s},
\label{eqn:efficiency}
\end{equation}
where $P_{\rm absorbed}$ is the total power trapped in the solar cell or solar thermal system, and $P_{\rm inc}$ is the total power incident on the device. The quantity $J_s$ is the incoming energy flux from sunlight, $J_e$ is the outgoing energy flux from photon emission, $J_r$ is the energy flux of sunlight reflected from the system, and $J_o$ are other thermal losses due to conduction, convection and non-radiative electron recombination. $J_s$ and $J_e$ are given by:
\begin{align}
\label{eqn:solareff}
J_{s}&=\Omega_c(N_s)\int_0^{\infty}d\omega\epsilon_a(\omega)I_s(\omega) \\
\label{eqn:reradiateeff}
J_{e}&=\pi\int_0^{\infty}d\omega\epsilon_e(\omega)I_{emit}(\omega)
\end{align}
where the effective absorptance $\epsilon_a(\omega)$ and emittance $\epsilon_e(\omega)$ are defined as:
\begin{align}
\label{eqn:ea}
\epsilon_{a}(\omega)&=\epsilon_t(\omega,\theta=\theta_t),\\
\label{eqn:ee}
\epsilon_{e}(\omega)&=\frac{1}{\pi}\int_{\phi=0}^{2\pi}\int_{\theta=0}^{\pi/2}\sin\theta~\cos\theta~\epsilon_t(\omega,\theta,\phi)~d\phi ~d\theta.
\end{align}
Here $\Omega_c(N_s)\approx N_s\cdot7\cdot10^{-5}$ rad is the solid angle spanned by the solar disk (at solar concentration of $N_s$ suns), $\epsilon_t$ is the transmission of angularly selective filter at different incident angles $(\theta,\phi)$ and frequency $f$ ($\epsilon_t=1$ if no angularly selective filter is applied). $\theta_t$ is the angle of incidence, $I_s$ is the solar spectrum, and $I_{emit}$ is the emission spectrum of solar cell or solar thermal system. Therefore, in order to maximize the solar trapping efficiency $\mu_I$, we need to minimize $J_e/J_s$ and $J_r/J_s$ (Eqn.~\ref{eqn:efficiency}). 

\begin{figure}[htbp]
\begin{center}
\includegraphics[width=3.2in]{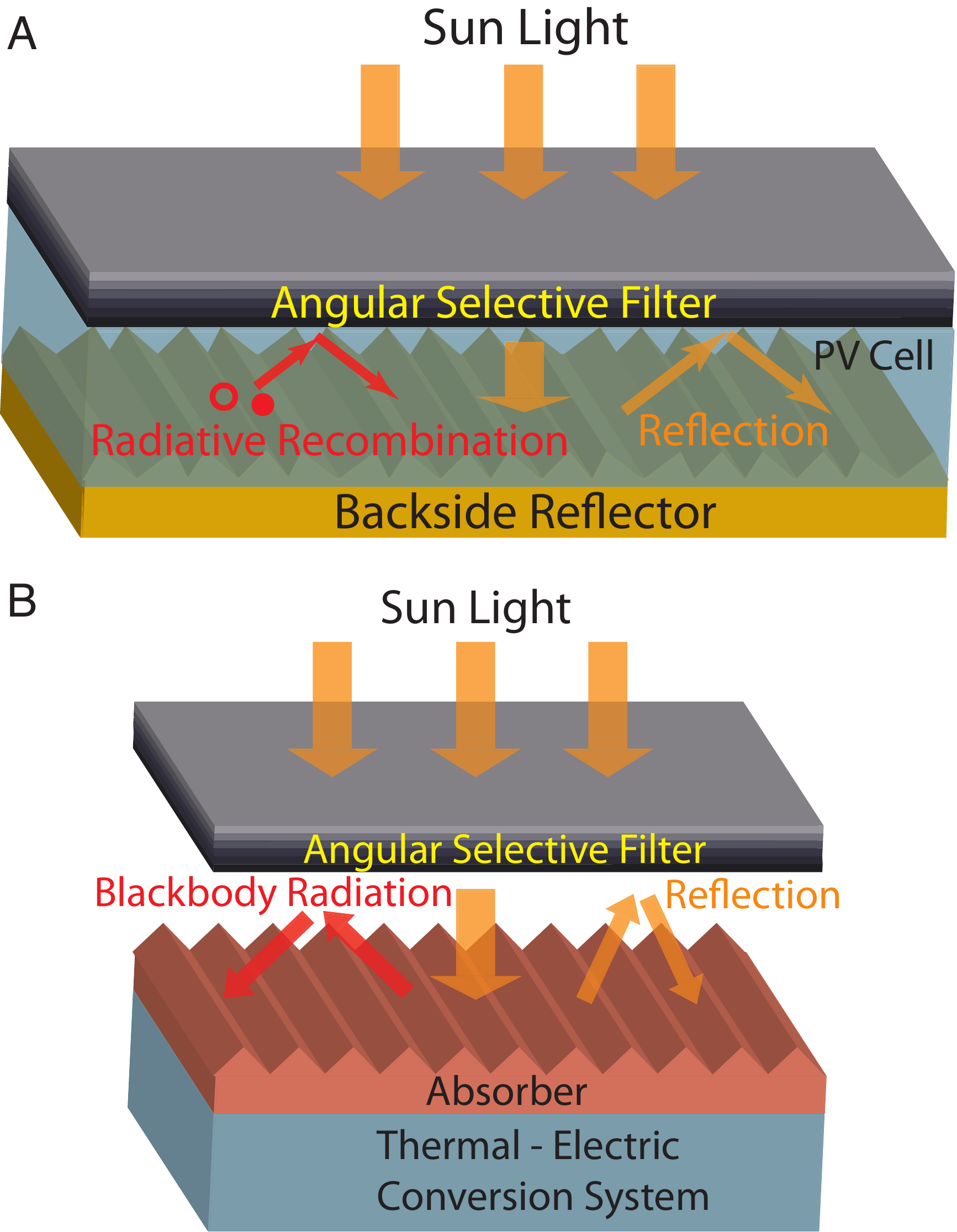}
\caption{\textbf{Application of Angular Selectivity in Energy Harvesting} (A) Solar Cell Application (B) Solar Thermal Application.}
\label{fig:angular_solar}
\end{center}
\end{figure}

When conventional solar cells or solar thermal absorbers are placed under direct sunlight, they receive light only from the solid angles spanned by the solar disk ($\Omega_{inc}=\theta_{c}(1)\approx6.8\cdot10^{-5}$ rad). On the other hand, they emit and reflect light isotropically ($\Omega_{emit}\approx2\pi$ rad). The large ratio between absorption and emission solid angles ($\Omega_{inc}/\Omega_{emit}\approx1.1\cdot10^{-5}$) results in an increase in photon entropy \cite{PSSA:PSSA200880460}, hence decreasing the efficiency of solar cells and solar thermal systems. There are two ways to increase $\Omega_{inc}/\Omega_{emit}$ \cite{peters2010angular}: The first way is to increase $\Omega_{inc}$ by applying a concentrator to the system \cite{fraas2010solar,weinstein2014optical,bett2003flatcon}. The second way is to decrease the angle of emission by applying angularly selective filters. The effect of such a filter is similar to using a concentrator, but can be made much thinner and can be easily incorporated into traditional solar cell modules. The angularly selective approach has been theoretically proposed several times \cite{2011SPIE.8124E..13K,0022-3727-38-13-014,kosten2014optical,Agrawal2009PhDT,Yablonovitch:82,doi:10.1117/12.921773,campbell1986limiting}. Up to now there have been only a few experimental realizations using narrow band angularly selective filters \cite{kosten2014experimental}. Methods of using broadband angularly selective filters in solar energy harvesting have been proposed theoretically \cite{Atwater_angular2013}, but remain challenging to fabricate. Using the implementation of the PhC angularly selective filter described in the previous section, we present here a new method to achieve broadband angular confinement (depicted in Fig.~\ref{fig:angular_solar}), and discuss its potential to improve the efficiency of current solar cells and solar thermal systems.

\subsubsection{Solar Cells}
In solar cells, the emission loss is mainly due to radiative recombination \cite{PhysRev.105.139}, and incomplete absorption (especially in thin film solar cells). A broadband angularly selective system can help mitigate losses from both of these causes, through the effects of photon-recycling (for radiative recombination) and light trapping (for reflected sunlight).

For solar cells with high radiative efficiency, such as GaAs or other direct-bandgap material solar cells, radiative recombination and emission is a major loss mechanism. In this case, photon recycling that reflects the radiated photons back into the solar cell using angularly selective filters can lead to enhancement in voltage and efficiency\cite{kosten2014optical,kosten2014experimental,Atwater_angular2013,hohn2012optimization}. In particular, a recent theoretical work from Atwater's group \cite{kosten2014optical} has predicted that by limiting the light emission angle, it is possible to achieve an enhancement of absolute efficiency by 8\% (without Auger recombination) and 5\% (with Auger recombination) for GaAs solar cells with ideal back reflector. In addition, it is also shown that the efficiency enhancement will increase significantly for thinner cells. Another recent work by Hohn \textit{et al.} \cite{hohn2014maximal} has compared the effects of the following three different angularly selective mechanisms in GaAs solar cells: 1) narrowband angularly selective filter with an optically flat mirror back reflector, 2) narrowband angularly selective filter with a diffusive back reflector, and 3) broadband angularly selective filter with a diffusive back reflector. The work shows theoretically that using a broadband angularly selective filter one can achieve a solar cell with the highest efficiency (41.3\%), compared with the Shockley-Queisser limit of 33.5\% \cite{miller2012strong}; and the lowest cell thickness (less than 100nm). However, such a broadband angularly selective filter was only hypothesized in both works mentioned above. 

On the other hand, for solar cells with low radiative efficiency, such as silicon solar cells, non-radiative processes like Auger recombination limit the voltage; hence photon recycling will have little effect on he efficiency\cite{marti1997,araujo1994}, and the light trapping effect is more significant. For solar cells with diffusive back reflector (Fig.~\ref{fig:angular_solar}A)\cite{muller2004}, it has been theoretically shown that limiting the emission angle in the \textit{solar spectrum} can reduce the optical escape cone and enhance the light trapping \cite{tiedje1984limiting,Yablonovitch:82,Atwater_angular2013,Kosten2015limiting}. Enhanced light trapping effect allows for better absorption of sunlight in a very thin cell (reducing material usage), and an increase in current, voltage and efficiency of the solar cell as well. A recent theoretical calculation from H. Atwater's group \cite{kosten2014optical, Kosten2015limiting} has shown that amorphous silicon hetero-junction with intrinsic thin layer (HIT) cells perform significantly better under angle restriction, with efficiency gains of approximately 1\% absolute achievable, with moderate angle restriction in parallel with a 50\% reduction in cell thickness.

\begin{figure}[htbp]
\begin{center}
\includegraphics[width=3in]{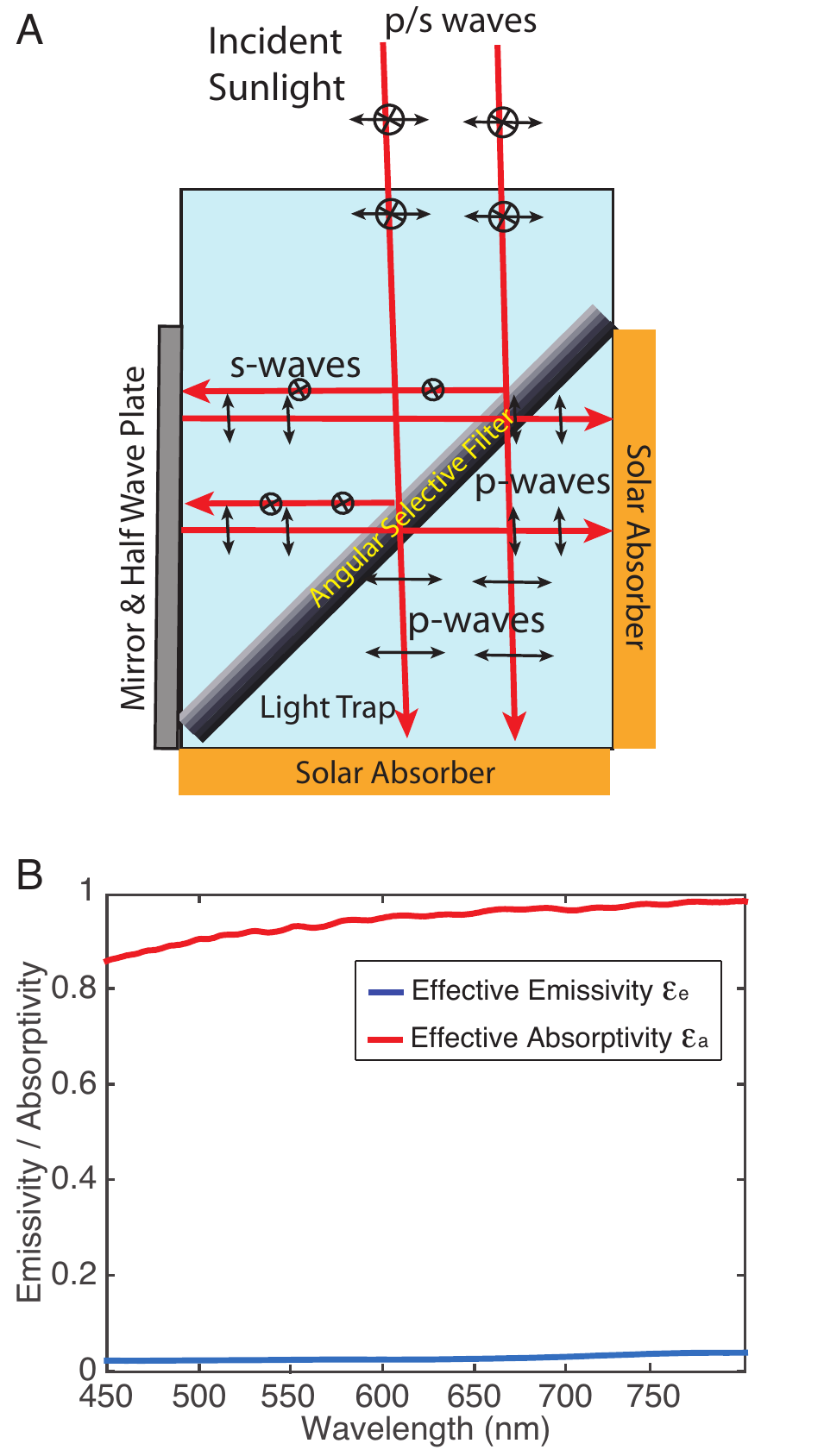}
\caption{(A) Experimental setup for polarization-independent angularly selective filter system. (B) Emittance and absorptance measurement. The red and blue curves are calculated through Eqn.~\ref{eqn:ea} and Eqn.~\ref{eqn:ee} with measured data of the sample used in our previous work \cite{Shen28032014}. 
}
\label{fig:both_polarization}
\end{center}
\end{figure}

Here we propose a new setup (Fig.~\ref{fig:both_polarization}) that achieves the broadband angular confinement using the angularly selective filter that Shen \textit{et al.} reported (Fig.~\ref{fig:brewster_prism})\cite{Shen28032014,air_compatible}. As illustrated in the inset of Fig.~\ref{fig:both_polarization}, although the PhC filter lets $p$-polarized light pass through only at the Brewster angle, one can use achromatic quarter-wave plate and a mirror to convert the randomly polarized sunlight into $p$-polarized light. Furthermore, the polarization restriction of the filter might also be lifted with the recent development of high efficiency polarizer \cite{arbabi2015dielectric,shen2014ultra}. We point out that since the PhC angularly selective window gets narrower as one increases the number of layers \cite{Shen28032014}, in principle the transparent angular window can be decreased further with more layers in the PhC filter. With a 1000-layer filter, the angular confinement (Fig.~2G in Ref.~\onlinecite{Shen28032014}) can be made less than 2$^{\circ}$. 

In conclusion, with the implementation of the broadband angularly selective filter, the enabled enhanced photon recycling and light trapping could allow for thinner solar cells with higher efficiency.

\subsubsection{Solar Thermal Systems}

Fig.~\ref{fig:angular_solar}B presents the schematic design of using the broadband angularly selective filter in solar thermal systems. Such broadband angularly selective filters can help increase the light trapping efficiency of current solar thermal systems through the following two ways:

Firstly, it can help eliminate the reflection loss from the absorber, or reduce $J_r/J_s$ in Eqn.~\ref{eqn:efficiency}. In order to effectively absorb sunlight, the solar thermal absorber is typically made up of metal with high thermal stability, such as tantalum or steel, so the surface reflection from the absorber is significant. For example, one of the most recently optimized solar thermal absorbers still exhibits a $13\%$ loss due to light reflection from the air-metal interface \cite{rinnerbauer2014,yeng2014global}. With the implementation of the broadband angularly selective filter, as illustrated in Fig.~\ref{fig:angular_solar}B and Fig.~\ref{fig:both_polarization}A, the reflected sunlight can be effectively trapped so that all the reflected light is recycled back to the absorber.

Secondly, it can help mitigate the photon emission loss from the absorber, or reduce $J_e/J_s$ in Eqn.~\ref{eqn:efficiency}. In traditional low temperature solar thermal systems, frequency selective absorbers are used to effectively separate the incident light and the emitted light \cite{lenert2014,Bermel:10,bermel2011tailoring}. However, there is a strong incentive to increase the absorber temperature since higher absorber temperature helps to achieve higher thermal-electric conversion (Eqn.~\ref{eqn:efficiency}). In the case of high temperature solar thermal systems, the overlap of $I_s$ and $I_{emit}$ in frequency is no longer negligible. A more detailed theoretical analysis by Blanco \textit{et al} \cite{blanco2004theoretical} suggests that in such a case one needs to rely on angular selection, although in this paper such broadband angularly selective materials are hypothetical.

The same setup we proposed in the solar cell section (Fig.~\ref{fig:both_polarization}) can also work for an angularly selective high-temperature solar thermal system. The transmittance of the angular confinement system at different incident angles is measured, and the calculated effective emittance $\epsilon_e$ and absorptance $\epsilon_a$ are plotted over the entire visible spectrum (Fig.~\ref{fig:both_polarization}). On average a factor of 47 enhancement is obtained for $\epsilon_a /\epsilon_e$ over the entire visible spectrum compared to the case without angularly selective filter ($\epsilon_a /\epsilon_e=1$). Such ratio can be enhanced even more by increasing the number of layers in the PhC slab, as explained in the previous work \cite{Shen28032014}. Furthermore, since the angularly selective filter does not need to be in a direct contact with the absorber, materials with high thermal stability are not required. 

\subsection{Transmitters and Detectors}
Among many detectors ranging from microscopes, cameras, radars, to telescopes, an important indicator used to describe their detection quality is the signal-to-noise ratio (SNR). In many cases, detectors are used to detect an object that usually spans only a small solid angle, while unwanted (noise) signals coming in from other directions can decrease the SNR. A similar argument can be applied to emitters: in many signal transmission applications, such as display screens, WiFi and radio stations, we only want the signal to propagate in certain directions. Signals propagating in other directions are either not wanted or present a waste of energy.

\begin{figure}[htbp]
\begin{center}
\includegraphics[width=3.2in]{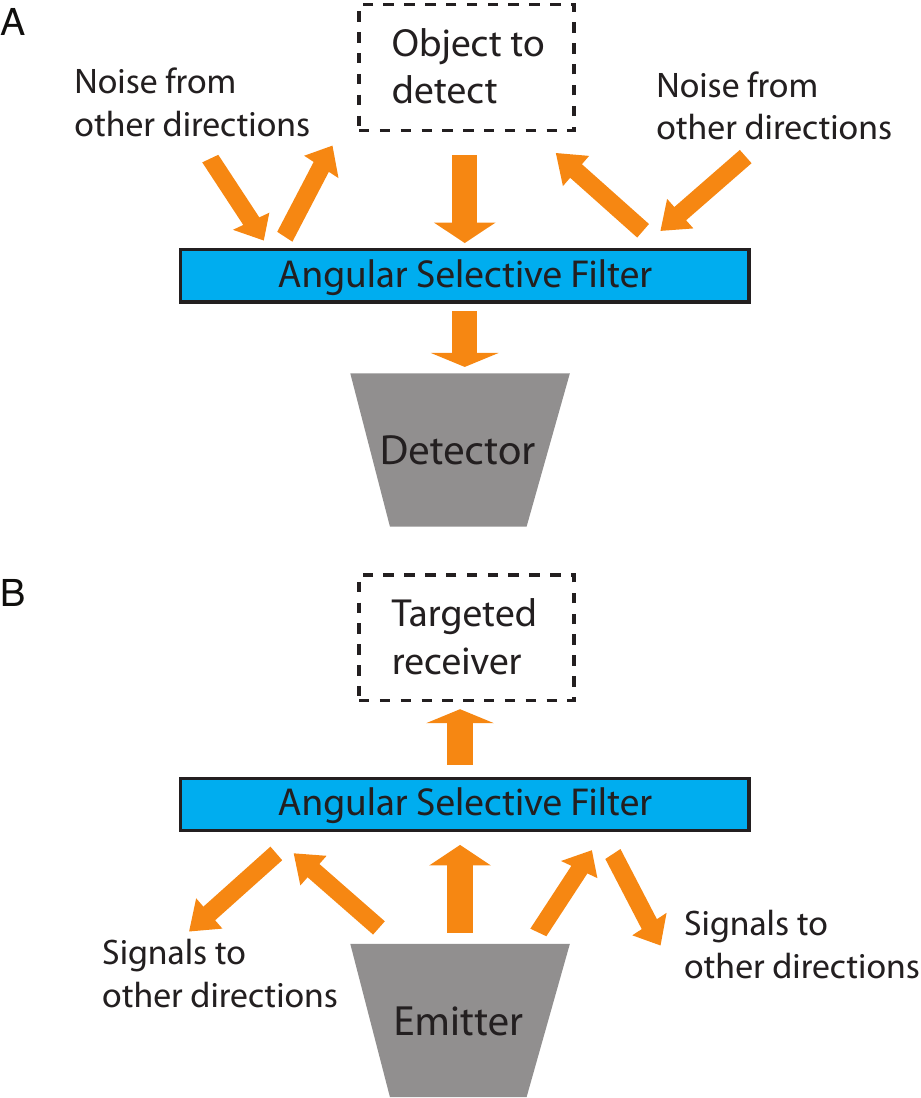}
\caption{(A) Illustration of enhancing signal/noise ratio by using angularly selective filter in front of an electromagnetic wave detector. (B) Illustration of enhancing directionality of light emitter using an angularly selective filter.}
\label{fig:detectors}
\end{center}
\end{figure}

In the past, various geometrical optic setups have been used to achieve angular selectivity and increase the SNR of the detectors or increase directionality of emitters. Such a setup can take up a lot of space and is sometimes the most expensive part of the whole system. Replacing the bulky and expensive traditional angularly selective system by the nanoscale, material-based angularly selective system (Fig.~\ref{fig:detectors}) has the potential to bring revolutionary changes to many important applications. In the following part of this section, we list a few major applications that are most suitable for the implementation of nanoscale broadband angularly selective system.

\begin{itemize}

\item \textbf{Transmitter Application: Privacy Screens}\\
As we discussed in section II, privacy screens are one of the earliest applications that motivated the study of angularly selective material systems. Today, privacy screens are widely used. In the United States, many healthcare workers are required by the Health Insurance Portability and Accountability Act of 1996 (HIPAA) to use privacy screens in the workplaces. Bankers and consultants in major firms are also required to use privacy screens when traveling or in public settings. 

The most widely used privacy screens today employ the traditional micro-louvre design described in Fig.~\ref{fig:microscale_optics}A. Since the absorptive micro-louvre also blocks part of the light even from normal incidence, such a filter typically reduces the brightness of the screen by 40\%-50\%. Furthermore, the periodic grating structure of the privacy film can interfere with the thin-film-transistor (TFT) layer inside the LCD screen and form Moire patterns on the screen. 

New broadband angularly selective filter based on nano photonics could lead to higher transmission at normal incidence, better angular selectivity and no Moire patterns.

\item \textbf{Detector Application: Radars, Telescopes}\\
Common detectors such as radars and telescopes are usually used to observe distant objects. Unfortunately unwanted signals coming from other directions can cause a reduction in SNR. In order to increase the SNR, typical radar systems use a steerable parabolic disk so that only light coming in at certain angle can be reflected to the receiving antenna (Fig.~1). Similarly, telescopes use lens/mirrors systems \cite{Refl_tele,swarup1991giant} or long detecting tube \cite{BICEP} to selectively filter out signals coming from unwanted directions. One common feature of this geometrical optics approach is that the angularly selective system needs to have the depth (size along the propagation direction of incident light) proportional to the detection aperture. Therefore, for those radars and telescopes that have large apertures, the angularly selective system would be very bulky, and sometimes even occupy the majority of the volume of the entire detecting system.

Here we propose a new design concept based on an angularly selective material system. As shown in Fig.~\ref{fig:detectors}, an angularly selective filter is added in front of the detector. Such implementation has been developed for narrow band detectors\cite{temelkuran2000photonic}, but not for broadband detectors yet. The angular sensitivity and bandwidth of this new kind of detection system can be made as good as the traditional detection systems (if not better); however, such an angularly selective filter can be made much thinner and independent of the aperture size, which has a potential to significantly reduce the system volume and might be valuable for many defense and commercial applications.

\end{itemize}

\section{Outlook}
Angularly selective devices based on geometrical optics have reached their limit due to the fact that their sizes need to be much larger than the operating wavelength. Nano photonics has opened many new opportunities in the development of broadband angularly selective materials, and is likely to have significant technological impact on this field. We have reviewed some of the most representative theoretical and experimental efforts in this field so far, namely using the micro-geometrical-optics, material birefringence, plasmonics, photonic crystals and metamaterials. 

Although this field has already seen rapid progress in the past few years, we believe that we have only seen the beginning of this important topic. Over the coming years, we expect the following major topics to be explored in order for nanoscale broadband angularly selective materials to be widely applicable to the applications discussed in the previous sections. 

\begin{itemize}
\item \textbf{Polarization dependence:}\\
The current realizations of the Brewster-angle-related methods described the sections II.B through II.E work only for one polarization. Generalization to both polarizations is a challenge, as the impedance-matching condition at the boundary depends on the polarization. This property applies generally to most PhCs and plasmonic structures irrespective of the dimensions. As a consequence, nano-photonic systems are typically polarization dependent.

However, this issue should not limit us from searching for broadband angularly selective systems that are independent of polarizations. For example, in Shen \textit{et al}'s work \cite{Shen28032014}, it is proposed that one can lift the restriction on polarization by using materials with nontrivial permeability $\mu\neq1$. During the past decades, it has been demonstrated that metamaterials have a good potential for achieving this. 

\item \textbf{Angularly selective reflection/scattering:}\\
An opposite but related effect is the angularly selective reflection and scattering of broadband light. These can be useful for applications such as radiative cooling devices \cite{raman2014passive} and transparent displays \cite{hsu2014transparent}. The impedance-matching-based approaches we reviewed are suitable for selective transmission but not for angularly selective reflection or scattering. To the best our knowledge, none of the existing technologies are capable of achieving such angularly selective reflection or scattering. Therefore, this topic remains to be explored.

\item \textbf{Tunable selective angle:}\\
As we discussed in the solar cell applications in section III, angularly selective materials can increase the efficiency of solar cells and solar thermal systems by a significant amount. But, in order to implement such materials, the transparent window would have to face the sun at all times. Therefore, when using the angularly selective materials with fixed selective angle, a solar tracker would be necessary. However, solar trackers require extra components and energy input, hence are only used in some very large scale solar fields. On the other hand, if one can develop a mechanism to dynamically tune the angle of  the angularly selective material system, it will solve the major problem  of solar tracking.

\end{itemize}

Following the rapid development of nanotechnology, the ability to control light at nanoscale has become an important topic and has attracted a tremendous amount of attention from both academia and industry. In particular, the field of broadband angular selectivity of light at the nanoscale has seen exciting progress in recent years. This new class of materials holds promising prospects for a host of applications, and thus deserves to be thoroughly explored both theoretically and experimentally.

In the coming years,we expect to see further development and more applications enabled by such technologies.

\begin{acknowledgments}
This work was partially supported by the Army Research Office through the ISN under Contract Nos.~W911NF-13-D0001. The fabrication part of the effort, as well as M.S. (reading and analysis of the manuscript) were supported by the MIT S3TEC Energy Research Frontier Center of the Department of Energy under Grant No. DE-SC0001299.
\end{acknowledgments}

\bibliography{Yichen_bib}

\end{document}